\documentclass[twocolumn,prl,showpacs,showkeys,runinaddress]{revtex4}%
\usepackage{amsfonts}
\usepackage{amsmath}
\usepackage{amssymb}
\usepackage{graphicx}%
\providecommand{\U}[1]{\protect\rule{.1in}{.1in}}
\begin{document}
\preprint{cond-mat/xxxx}
\title{Contact-less measurements of Shubnikov-de Haas oscillations in the
magnetically ordered state of CeAgSb$_{2}$ and SmAgSb$_{2}$ single crystals}
\author{Ruslan Prozorov}
\email{prozorov@ameslab.gov}
\affiliation{Ames Laboratory and Department of Physics \& Astronomy, Iowa State University,
Ames, Iowa 50011}
\author{Matthew D. Vannette}
\affiliation{Ames Laboratory and Department of Physics \& Astronomy, Iowa State University,
Ames, Iowa 50011}
\author{German D. Samolyuk}
\affiliation{Ames Laboratory and Department of Physics \& Astronomy, Iowa State University,
Ames, Iowa 50011}
\author{Stephanie A. Law}
\affiliation{Ames Laboratory and Department of Physics \& Astronomy, Iowa State University,
Ames, Iowa 50011}
\author{Sergey L. Bud'ko}
\affiliation{Ames Laboratory and Department of Physics \& Astronomy, Iowa State University,
Ames, Iowa 50011}
\author{Paul C. Canfield}
\affiliation{Ames Laboratory and Department of Physics \& Astronomy, Iowa State University,
Ames, Iowa 50011}

\begin{abstract}
Shubnikov - de Haas oscillations were measured in single crystals of highly
metallic antiferromagnetic SmAgSb$_{2}$ and ferromagnetic CeAgSb$_{2}$ using a
tunnel diode resonator. Resistivity oscillations as a function of applied
magnetic field were observed via measurements of skin depth variation. The
effective resolution of $\Delta\rho\simeq20$ p$\Omega$ allows a detailed study
of the SdH spectra as a function of temperature. The effects of the Sm long -
range magnetic ordering as well as its electronic structure ($4f$-electrons)
on the Fermi surface topology is discussed.

\end{abstract}
\date{22 May 2006}
\maketitle

\section{Introduction}

Measurements of quantum oscillations in the resistivity (Shubnikov - de Haas
effect (SdH)) and in the magnetization (de Haas - van Alphen effect (dHvA))
are among the primary experimental techniques to study the geometry of Fermi
surfaces in metals \cite{shoenberg,kartsovnik}. Although both originate from
the basic physics of Landau quantization of electron orbits in a magnetic
field, the coupling of electron motion to resistivity is different from that
of magnetization. Electron transport depends on the density of states and the
scattering rates, both modulated by the Landau quantization. Within a standard
theory \cite{shoenberg} for a three-dimensional Fermi surface the amplitude of
the $r^{th}$ harmonic oscillatory part of magnetization is given by
\cite{kartsovnik},
\begin{equation}
M_{r}\propto\frac{S_{extr}B^{1/2}}{m_{c}\left\vert S"\right\vert _{extr}%
^{1/2}}R_{T}\left(  r\right)  R_{D}\left(  r\right)  R_{S}\left(  r\right)
\label{dHvA}%
\end{equation}
where $m_{c}$ is the cyclotron mass, $S_{extr}$ is the extremal cross-section
of the Fermi surface (FS), $\left\vert S"\right\vert _{extr}=\left(
\partial^{2}S/\partial p_{B}^{2}\right)  _{extr}$ is a measure of the FS
curvature along $\mathbf{B}$ at the extremal cross-section, and the damping
factors $R_{T}$, $R_{D}$, and $R_{S}$ are caused by finite temperature,
scattering and Zeeman splitting, respectively. The corresponding fundamental
frequency is%
\begin{equation}
f=\frac{S_{extr}}{he} \label{fext}%
\end{equation}

On the other hand, the amplitude of Shubnikov - de Haas oscillations in
electrical resistivity is proportional to
\begin{equation}
\alpha_{r}\propto\frac{m_{c}S_{extr}B^{1/2}}{\left\vert S"\right\vert
_{extr}^{1/2}}R_{T}\left(  r\right)  R_{D}\left(  r\right)  R_{S}\left(
r\right)  \label{SdH}%
\end{equation}
with the same damping factors as for dHvA effect. Typically, oscillations of
resistivity are more difficult to measure, especially in highly conducting
samples. The problem is worse for small single crystals where attaching the
contacts is not an easy task. It should be noted that the SdH amplitude
dependence on the $m_{c}$ is reciprocal to the dHvA amplitudes: $\alpha
_{r}/M_{r}\propto m_{c}^{2}$, so for small $m_{c}$, the relative amplitude of
the SdH oscillations is suppressed.

The alternative to direct transport measurements is to measure the skin depth.
This paper describes the use of highly sensitive tunnel-diode oscillator
technique for the quantitative study of quantum oscillations in metallic
samples. The technique is especially useful for small samples when attachment
of contacts is difficult. Detailed analysis of the raw data is presented.

\section{The technique}

An oscillating magnetic field of frequency $f$ penetrates a metallic sample,
decaying in a typical length $\delta$, the skin depth. In the local limit,
where $E=\rho j$, $\delta$ is given by the normal skin effect expression
\cite{jackson},
\begin{equation}
\delta=\frac{c}{2\pi}\sqrt{\frac{\rho}{\mu f}} \label{delta}%
\end{equation}
where $\rho$ is resistivity and $\mu$ is magnetic permeability. Unfortunately,
typical AC magnetometers are not sensitive enough to measure this depth with
sufficient precision. In addition, for typical low-frequency magnetometers (up
to kHz frequencies), skin depth is too large compared to the small crystal
size. For higher-frequency techniques, such as microwave or infrared
reflectivity and absorption, the anomalous skin effect and possible dynamic
effects (e.g. magnetic relaxation) take place. Although the anomalous skin
effect was originally used to study Fermi surfaces \cite{pippard}, in the
extreme anomalous limit the skin effect is no longer determined only by the
resistivity, but also by the electronic mean free path, $\ell$. The skin depth
then depends on a product $\rho\ell$, which is independent of the scattering
rate and cannot be used for contact-less measurements of resistivity.

In this work, a self-resonating, tunnel diode driven LC circuit was used for
precise measurements of the effect of temperature and magnetic field on the
skin depth of CeAgSb$_{2}$ and SmAgSb$_{2}$. The circuit had a natural
oscillation frequency of $\simeq10~$MHz. Originally, this technique was
developed to measure London penetration depth in small superconducting
crystals, where its great sensitivity was a key factor. Details of the
apparatus and calibration are given elsewhere \cite{prozorov,prozorov2}.
Tunnel diode resonators have been used for qualitative studies of the skin
depth oscillations in pulsed magnetic fields in organic superconductors with
relatively large normal-state resistivities ( $>200$ $\mu%
%TCIMACRO{\unit{\U{3a9}}}%
%BeginExpansion
\operatorname{\Omega }%
%EndExpansion
\cdot%
%TCIMACRO{\unit{cm}}%
%BeginExpansion
\operatorname{cm}%
%EndExpansion
$) \cite{choi,coffey}. Although little quantitative analysis was reported,
these studies demonstrate the effectiveness of the technique.

When a metallic sample is inserted into a coil which is part of an LC circuit
the resonant frequency changes from $f_{0}$ to $f$. This change is caused by
two effects: the diamagnetic skin effect, which screens the alternating
magnetic field from the bulk of the sample and the magnetic permeability $\mu$
of the sample. The measured frequency shift $\Delta f=f\left(  T,H\right)
-f_{0}$ is given by the total dynamic magnetic susceptibility of the sample,
$\chi=dM/dH$ \cite{prozorov},%
\begin{equation}
\frac{\Delta f}{\Delta f_{0}}=1-\frac{\mu\delta}{2R}\tanh\left(  \frac{2R}%
{\mu\delta}\right)  \label{df}%
\end{equation}
where $R$ is the effective sample dimension (for an infinite slab of width
$w$, $R=w/2$) and skin depth $\delta$ is given by Eq.(\ref{delta}). The
maximal expulsion frequency, $\Delta f_{0}$, (frequency shift in the limit of
a perfect conductor with $\delta=0$) depends only on the geometry of the
sample and parameters of empty resonator and is given by
\begin{equation}
\Delta f_{0}=\frac{fV}{2V_{0}\left(  1-N\right)  } \label{df0}%
\end{equation}
where $V$ is the sample volume, $V_{0}$ is the effective coil volume and $N$
is the demagnetization factor \cite{prozorov}. In deriving Eq.(\ref{df}) it is
assumed that $\Delta f_{0}\ll f_{0}$, which is always the case in our
frequency and sample size range (the base frequency is approximately
$f_{0}=10^{7}$ Hz, whereas typical $\Delta f_{0}=10^{4}$ Hz). The factor $\mu$
in Eq.(\ref{df}) comes from the direct magnetic contribution of magnetic
permeability in the skin layer via $B=\mu H$, which enters the total magnetic
moment after integration over the sample volume. A similar effect was
discussed before for superconductors, \cite{cooper}. However, the direct
contribution of magnetic permeability in the samples under study is minor.
This is because we obtain useful information only well below the ordering
temperature and in high magnetic fields where $\mu\approx1$. The oscillating
part of the permeability is small due to very small excitation field,
$H_{ac}\approx20$ mOe. In addition, magnetic field only penetrates the skin
depth layer and its direct contribution to the total susceptibility is
attenuated by the factor of $\left(  \delta/R\right)  $ compared to the skin
effect contribution. In other magnetic systems with large polarizability or in
the vicinity of the magnetic ordering temperature, direct magnetic
contribution could be important, but this is not relevant to the present work.

Obviously, the described technique is sensitive only as long as the skin depth
is less than the effective sample size. For larger skin depth (larger
resistivity), the sample becomes transparent to an oscillating magnetic field.
For example, in our particular setup in actual units,%
\begin{equation}
\delta\left[
%TCIMACRO{\unit{\U{3bc}m}}%
%BeginExpansion
\operatorname{\mu m}%
%EndExpansion
\right]  =15\sqrt{\rho\left[  \mu%
%TCIMACRO{\unit{\U{3a9}}}%
%BeginExpansion
\operatorname{\Omega }%
%EndExpansion
\cdot%
%TCIMACRO{\unit{cm}}%
%BeginExpansion
\operatorname{cm}%
%EndExpansion
\right]  } \label{dum}%
\end{equation}
With small crystals of typical size of $R=500$ $\mu$m, we can roughly estimate
the upper limit where the described technique is sensitive. By equating
$\delta_{\max}=R$, we obtain $\rho_{\max}=\left(  500/15\right)  ^{2}%
\approx1000$ $\mu%
%TCIMACRO{\unit{\U{3a9}}}%
%BeginExpansion
\operatorname{\Omega }%
%EndExpansion
\cdot%
%TCIMACRO{\unit{cm}}%
%BeginExpansion
\operatorname{cm}%
%EndExpansion
$. Measurements on samples with larger resistivities are possible, but require
larger samples or higher frequencies (both of which are possible). At the
opposite extreme for samples with very small resistivities, the limitation is
the anomalous skin effect, which becomes relevant when skin depth becomes
smaller than the electronic mean free path, $\ell$, and the local version of
the Ohm's law is no longer valid. We can roughly estimate the lower threshold
of resistivity below which anomalous skin effect takes place by using a Drude
approximation for which the mean free path is given by%
\begin{equation}
\ell=\frac{\left(  3\pi^{2}\right)  ^{1/3}\hbar}{n^{2/3}e^{2}\rho} \label{mfp}%
\end{equation}
where $n$ is the electron density and $e$ is the electron charge. Therefore,
using Eq.(\ref{delta}) we obtain $\rho_{\min}$ at which $\ell=\delta$,%
\begin{equation}
\rho_{\min}=\left(  \frac{\mu f\pi^{7/3}3^{2/3}\hbar^{2}}{n^{4/3}e^{4}%
}\right)  ^{\frac{1}{3}}\approx8.6\times\frac{\left(  \mu f\right)  ^{1/3}%
}{n^{4/9}} \label{rmin}%
\end{equation}
For a typical nonmagnetic metal with $n=5\times10^{28}$ m$^{-3}$ and our
frequency, $f=10^{7}$ Hz, we obtain $\rho_{a}\approx0.03$ $\mu\Omega\cdot$ cm.
Therefore, our method allows direct quantitative study of (contact-less)
resistivity in small crystals approximately in the range of $0.03-1000$
$\mu\Omega\cdot$ cm, which covers most metallic materials of interest.
Furthermore, the difficulty of measuring small highly conducting samples by
conventional means turns to be an advantage for the data analysis. In the
regime of good metals, the data is significantly simplified, because the
$\tanh\left(  R/\delta\right)  $ term in Eq.(\ref{df}) becomes relevant only
for $\delta\geq0.2R$. In the samples discussed below, the resistivity varies
between $0.1-10$ $\mu%
%TCIMACRO{\unit{\U{3a9}}}%
%BeginExpansion
\operatorname{\Omega }%
%EndExpansion
\cdot%
%TCIMACRO{\unit{cm}}%
%BeginExpansion
\operatorname{cm}%
%EndExpansion
$, for which $\delta/R=0.01-0.1$ and therefore, we can simply use the linear
version of Eq.(\ref{df}).
\begin{equation}
\frac{\Delta f}{\Delta f_{0}}=1-\frac{\mu\delta}{2R} \label{dflinearized}%
\end{equation}
It should be noted that it is still straightforward to invert full equation
Eq.(\ref{df}) to obtain the skin depth numerically.

In the measurements reported below, slab - shaped samples were positioned
inside the resonant coil on a sapphire sample holder. The crystallographic
$c-$ axis was parallel to both the AC excitation and DC magnetic fields. In
this arrangement, screening currents flowed in the $ab-$ plane and therefore
the in-plane component of the resistivity was measured. A typical run was from
$0$ to $9$ Tesla DC field and back to $0$ T with about $10000$ data points
taken in each leg. Various ramp rates where used with the slowest being $3.5$ G/s.

\section{Samples}

In this work, the skin depth was measured in two isostructural tetragonal
intermetallic compounds: CeAgSb$_{2}$ and SmAgSb$_{2}$ single crystals. Both
materials (and related compounds) were studied before
\cite{brylak,sologub,myers1,myers2,jobiliong}, though with much more attention
to CeAgSb$_{2}$ which undergoes a ferromagnetic transition along the $c-$ axis
below $T_{c}=9.8$ K. SmAgSb$_{2}$ is antiferromagnetic with $T_{N}=9.5$ K.
Transport and magnetic properties, including direct measurements of dHvA and
SdH effects were reported previously \cite{myers1,myers2,jobiliong}. Whereas
SdH in CeAgSb$_{2}$ showed a single dominant frequency at $0.25$ MG, in
SmAgSb$_{2}$ the behavior is much more complex with several frequencies and
significant difference between dHvA and SdH spectra. One difference between
CeAgSb$_{2}$ and SmAgSb$_{2}$ is substantially larger scattering rates in the
former (as evident from $RRR$). Also, as we show below, small changes in the
Fermi level lead to significant changes of the Fermi surface structure and
appearance of new extremal orbits.

\begin{figure}
[ptbh]
\begin{center}
\includegraphics[
height=3.508cm,
width=8.0792cm
]%
{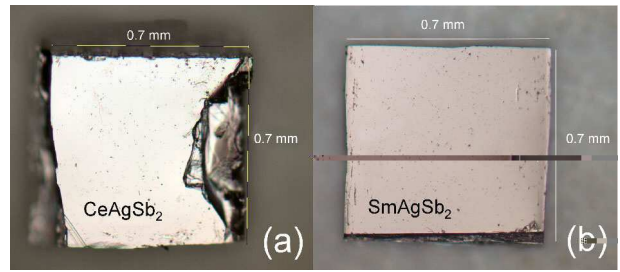}%
\caption{Optical images of studied crystals. (a) CeAgSb2 and (b) SmAgSb2. Each
crystal was $0.7\times0.7\times0.3$ mm$^{3}$.}%
\label{samples}%
\end{center}
\end{figure}

Single crystals of CeAgSb$_{2}$ and SmAgSb$_{2}$ were grown out of Sb flux
\cite{canfield,myers1}. The starting materials were placed in an alumina
crucible and sealed under vacuum in a quartz ampule, heated to 1150
%TCIMACRO{\U{b0}}%
%BeginExpansion
${{}^\circ}$%
%EndExpansion
C, and then cooled to 670
%TCIMACRO{\U{b0}}%
%BeginExpansion
${{}^\circ}$%
%EndExpansion
C over 120 hours. The best crystals with the cleanest surface were selected
and cut with a blade along the sides for the measurements. The samples are
shown in Fig. \ref{samples}. The c-axis was perpendicular to the largest face.
The residual resistivity ratio, $RRR=\rho\left(  300%
%TCIMACRO{\unit{K}}%
%BeginExpansion
\operatorname{K}%
%EndExpansion
\right)  /\rho\left(  2%
%TCIMACRO{\unit{K}}%
%BeginExpansion
\operatorname{K}%
%EndExpansion
\right)  $ was determined by standard four-point measurements. For
SmAgSb$_{2}$ we obtained $RRR=200$ and for CeAgSb$_{2}$ $RRR=70$. The
transport measurements were performed on bar-shaped samples from the same
batches that were used for resonator measurements, shown in Fig.\ref{samples}.%

%TCIMACRO{\FRAME{ftbphFU}{8.0792cm}{3.508cm}{0pt}{\Qcb{Optical images of
%studied crystals. (a) CeAgSb2 and (b) SmAgSb2. Each crystal was $0.7\times
%0.7\times0.3$ mm$^{3}$.}}{\Qlb{samples}}{samples_small.eps}%
%{\special{ language "Scientific Word";  type "GRAPHIC";
%maintain-aspect-ratio TRUE;  display "ICON";  valid_file "F";
%width 8.0792cm;  height 3.508cm;  depth 0pt;  original-width 3.1531in;
%original-height 1.3534in;  cropleft "0";  croptop "1";  cropright "1";
%cropbottom "0";  filename 'samples_small.eps';file-properties "XNPEU";}}}%
%BeginExpansion
%EndExpansion

\section{Results}

\begin{figure}
[ptbh]
\begin{center}
\includegraphics[
height=2.802in,
width=3.5561in
]%
{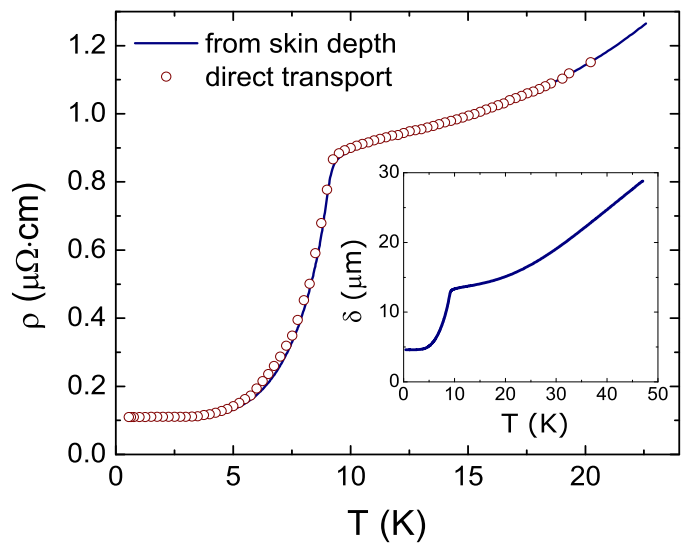}%
\caption{Resitivity versus temperature for SmAgSb$_{2}$ single crystal
measured directly (open symbols) and inferred from the skin depth measurements
(solid line). The inset shows variation of the skin depth with temperature.}%
\label{SmRvsT}%
\end{center}
\end{figure}

Figure \ref{SmRvsT} shows resistivity versus temperature obtained from the
frequency shift by using Eq.(\ref{df}) and (\ref{delta}). The skin depth is
shown in the inset. The high-temperature (at $T=20$ K) resistivity data was
used to normalize the calibration constant, $\Delta f_{0}$. Direct calibration
with known parameters of the oscillator produce very similar temperature
dependence, but the offset (residual resistivity) cannot be obtained from this
technique, because it only measured the relative frequency change. However,
this is only beneficial for the present study, as we are only interested in
changes of resisivity in a magnetic field and are not sensitive to static
resistivity background.%

%TCIMACRO{\FRAME{ftbphFU}{3.5561in}{2.802in}{0pt}{\Qcb{Resitivity versus
%temperature for SmAgSb$_{2}$ single crystal measured directly (open symbols)
%and inferred from the skin depth measurements (solid line). The inset shows
%variation of the skin depth with temperature.}}{\Qlb{SmRvsT}}{smrvst.eps}%
%{\special{ language "Scientific Word";  type "GRAPHIC";
%maintain-aspect-ratio TRUE;  display "ICON";  valid_file "F";
%width 3.5561in;  height 2.802in;  depth 0pt;  original-width 3.5284in;
%original-height 2.7743in;  cropleft "0";  croptop "1";  cropright "1";
%cropbottom "0";  filename 'SmRvsT.eps';file-properties "XNPEU";}}}%
%BeginExpansion
%EndExpansion

As expected, due to an increase of spin-disorder and phonon scattering, the
resistivity sharply decreases below $T_{N}$. The reconstructed resistivity
(solid line) is compared to direct four-point measurements performed on the
sample from the same batch. Evidently, the quantitative agreement is good.
Similarly, a good agreement was obtained for $\rho\left(  T\right)  $ measured
in various applied magnetic fields where the magneto-resistance followed a
Kohler rule as described in detail in Ref.\cite{myers1}.

\begin{figure}
[ptbh]
\begin{center}
\includegraphics[
height=4.1684in,
width=3.3806in
]%
{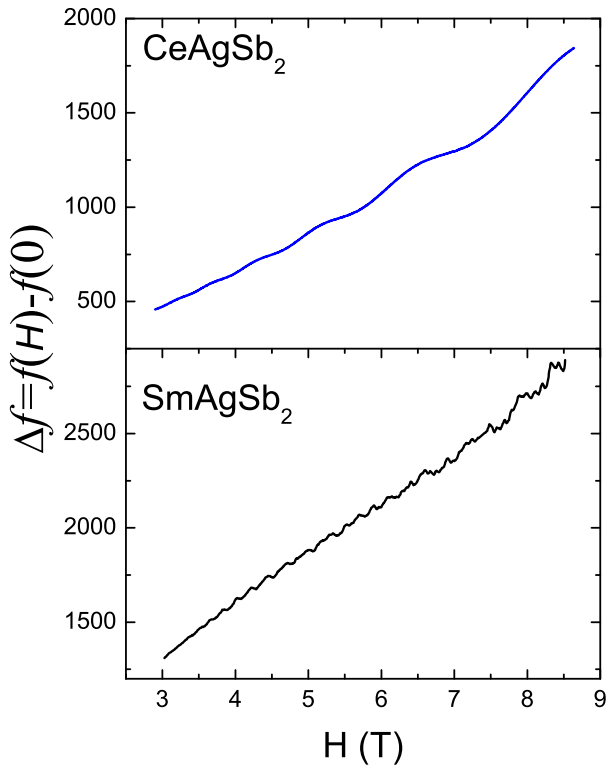}%
\caption{Relative frequency shift in a magnetic field for CeAgSb$_{2}$ (top)
and SmAgSb$_{2}$ (bottom) at 1.8 K.}%
\label{raw}%
\end{center}
\end{figure}

The oscillations of resistivity were obtained from the measured oscillations
in the skin depth, which in turn were obtained from the oscillating frequency
shift. An example of the raw data for both, CeAgSb$_{2}$ and SmAgSb$_{2}$ are
shown in Fig. \ref{raw} in the top and bottom panels, respectively.%

%TCIMACRO{\FRAME{ftbphFU}{3.3806in}{4.1684in}{0pt}{\Qcb{Relative frequency
%shift in a magnetic field for CeAgSb$_{2}$ (top) and SmAgSb$_{2}$ (bottom) at
%1.8 K.}}{\Qlb{raw}}{bkgrnd.eps}{\special{ language "Scientific Word";
%type "GRAPHIC";  maintain-aspect-ratio TRUE;  display "ICON";
%valid_file "F";  width 3.3806in;  height 4.1684in;  depth 0pt;
%original-width 3.3338in;  original-height 4.1182in;  cropleft "0";
%croptop "1";  cropright "1";  cropbottom "0";
%filename 'BKGRND.eps';file-properties "XNPEU";}}}%
%BeginExpansion
\begin{figure}
[ptbh]
\begin{center}
\includegraphics[
height=9.7838cm,
width=7.7607cm
]%
{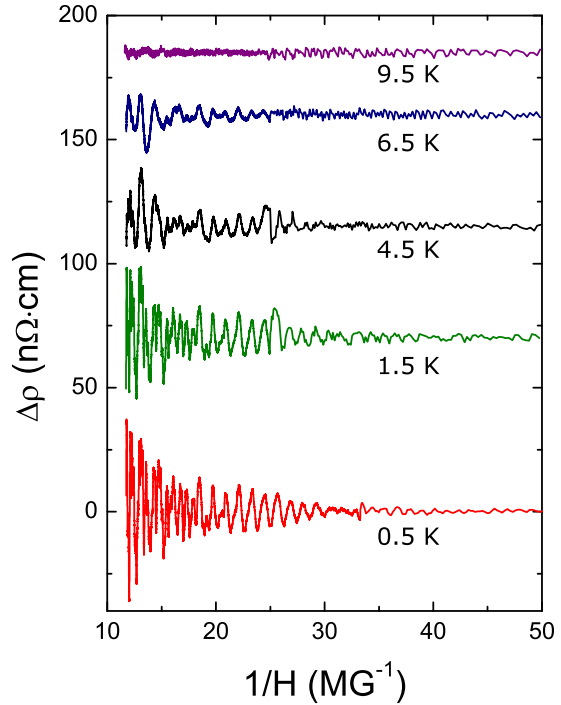}%
\caption{Oscillations of the reconstructed resistivity (note the resistivity
scale (the curves are offset for clarity)) versus reciprocal magnetic field in
SmAgSb$_{2}$ single crystal measured at several temperatures below $T_{c}$.}%
\label{SmSdH}%
\end{center}
\end{figure}

%EndExpansion
For SmAgSb$_{2}$ the resistivity oscillations reconstructed from the frequency
shift are shown as a function of $H^{-1}$ in Fig. (\ref{SmSdH}). A smooth
$\rho\left(  H\right)  $ background was subtracted using non-oscillating
piecewise cubic hermite interpolating polynomial algorythm in Matlab. The
amplitude of the oscillations diminishes for temperatures approaching the
Ne\'{e}l temperature from below, as anticipated from the increasing spin
disorder scattering evident from Fig.\ref{SmRvsT}.%

%TCIMACRO{\FRAME{ftbphFU}{7.7607cm}{9.7838cm}{0pt}{\Qcb{Oscillations of the
%reconstructed resistivity (note the resistivity scale (the curves are offset
%for clarity)) versus reciprocal magnetic field in SmAgSb$_{2}$ single crystal
%measured at several temperatures below $T_{c}$.}}{\Qlb{SmSdH}}{sdhrho.eps}%
%{\special{ language "Scientific Word";  type "GRAPHIC";
%maintain-aspect-ratio TRUE;  display "ICON";  valid_file "F";
%width 7.7607cm;  height 9.7838cm;  depth 0pt;  original-width 3.0277in;
%original-height 3.8242in;  cropleft "0";  croptop "1";  cropright "1";
%cropbottom "0";  filename 'SdHrho.eps';file-properties "XNPEU";}}}%
%BeginExpansion
%EndExpansion

\begin{figure}
[ptbh]
\begin{center}
\includegraphics[
height=9.7157cm,
width=7.5827cm
]%
{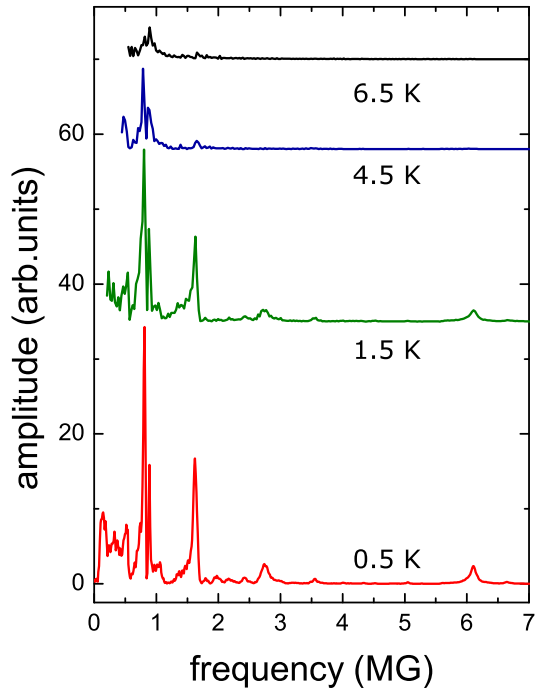}%
\caption{Power spectra of Shubnikov-de Haas oscillations obtained from the
data shown in Fig.(\ref{SmSdH}). The curves are offset for clarity.}%
\label{SmPSdiffT}%
\end{center}
\end{figure}

Power spectra were obtained from the oscillations by using a Fourier
transformation. The result is shown in Fig.(\ref{SmPSdiffT}). At a first
glance, the spectra are quite similar to those reported from the previous
direct measurements of the SdH oscillations \cite{myers2}. However, closer
inspection reveals additional details, most likely due to higher sensitivity
of our measurements.%

%TCIMACRO{\FRAME{ftbphFU}{7.5827cm}{9.7157cm}{0pt}{\Qcb{Power spectra of
%Shubnikov-de Haas oscillations obtained from the data shown in
%Fig.(\ref{SmSdH}). The curves are offset for clarity.}}{\Qlb{SmPSdiffT}%
%}{smpsvst.eps}{\special{ language "Scientific Word";  type "GRAPHIC";
%maintain-aspect-ratio TRUE;  display "ICON";  valid_file "F";
%width 7.5827cm;  height 9.7157cm;  depth 0pt;  original-width 2.9577in;
%original-height 3.7965in;  cropleft "0";  croptop "1";  cropright "1";
%cropbottom "0";  filename 'SmPSvsT.eps';file-properties "XNPEU";}}}%
%BeginExpansion
%EndExpansion

\begin{figure}
[ptbh]
\begin{center}
\includegraphics[
height=7.6069cm,
width=8.9952cm
]%
{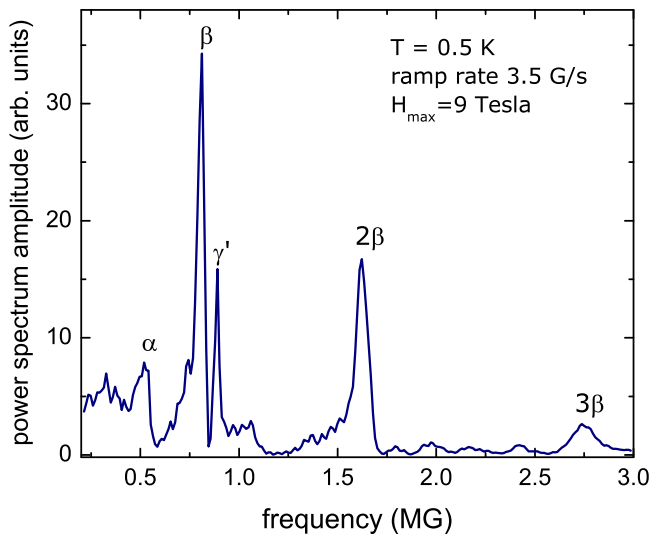}%
\caption{Low-frequency portion of the power spectra of Shubnikov-de Haas
oscillations measured in SmAgSb$_{2}$ at $T=0.5$ K.}%
\label{SmPST05}%
\end{center}
\end{figure}

Figure (\ref{SmPST05}) shows low-frequency detailed spectrum of oscillations
obtained at $T=0.5$ K. In addition to previously observed main $\beta$ peak,
there is a sharp adjacent peak at 0.9 MG, which was apparently unresolved in
previous direct measurements, at least down to 1.8 K. This new peak is not a
secondary combination of $\alpha$ and $\beta$ peaks and we attribute it to the
$\gamma^{\prime}$ orbit shown below in Fig. \ref{banda} (a).%

%TCIMACRO{\FRAME{ftbphFU}{8.9952cm}{7.6069cm}{0pt}{\Qcb{Low-frequency portion
%of the power spectra of Shubnikov-de Haas oscillations measured in
%SmAgSb$_{2}$ at $T=0.5$ K.}}{\Qlb{SmPST05}}{smps05.eps}%
%{\special{ language "Scientific Word";  type "GRAPHIC";
%maintain-aspect-ratio TRUE;  display "ICON";  valid_file "F";
%width 8.9952cm;  height 7.6069cm;  depth 0pt;  original-width 3.5137in;
%original-height 2.9672in;  cropleft "0";  croptop "1";  cropright "1";
%cropbottom "0";  filename 'SmPS05.eps';file-properties "XNPEU";}}}%
%BeginExpansion
%EndExpansion

\begin{figure}
[ptbh]
\begin{center}
\includegraphics[
height=7.6465cm,
width=9.0325cm
]%
{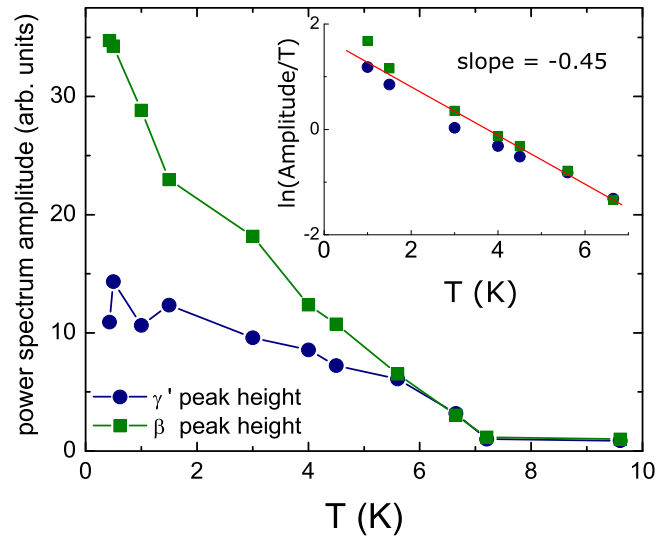}%
\caption{Temperature dependence of $\beta$ and $\gamma^{\prime}$ peaks power
spectrum amplitudes in SmAgSb$_{2}$. Inset: $\log$ of Fourier transform
amplitude over $T$ - used in calculation of the effective mass.}%
\label{SmAmpT}%
\end{center}
\end{figure}

Since our measurements are quantitative, it is possible to plot temperature
dependence of the observed peaks. Figure (\ref{SmAmpT}) demonstrates that
these dependencies are quite different. The $\beta$ peak is suppressed much
more rapidly. Both peaks vanish as the Ne\'{e}l temperature is approached.%

%TCIMACRO{\FRAME{ftbphFU}{9.0325cm}{7.6465cm}{0pt}{\Qcb{Temperature dependence
%of $\beta$ and $\gamma^{\prime}$ peaks power spectrum amplitudes in
%SmAgSb$_{2}$. Inset: $\log$ of Fourier transform amplitude over $T$ - used in
%calculation of the effective mass.}}{\Qlb{SmAmpT}}{smapmtdep.eps}%
%{\special{ language "Scientific Word";  type "GRAPHIC";
%maintain-aspect-ratio TRUE;  display "ICON";  valid_file "F";
%width 9.0325cm;  height 7.6465cm;  depth 0pt;  original-width 3.5284in;
%original-height 2.9827in;  cropleft "0";  croptop "1";  cropright "1";
%cropbottom "0";  filename 'SmApmTdep.eps';file-properties "XNPEU";}}}%
%BeginExpansion
%EndExpansion

\begin{figure}
[ptbh]
\begin{center}
\includegraphics[
height=9.7838cm,
width=7.7607cm
]%
{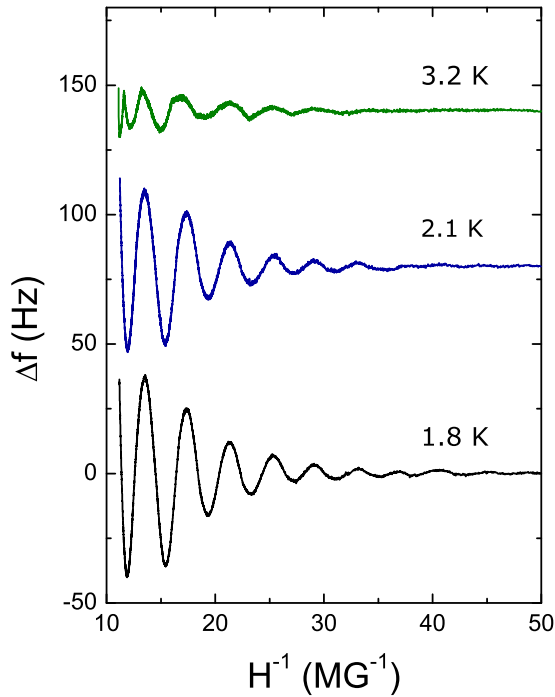}%
\caption{Shubnikov - de Haas oscillations (raw data in terms of the resonant
frequency shift) in single crystal CeAgSb$_{2}$ measured at indicated
temperatures. The curves are offset for clarity.}%
\label{CeOsc}%
\end{center}
\end{figure}

In principle, we cannot rule out the influence of de Haas - van Alphen
oscillations of magnetization (via $\mu$ term in Eq. (\ref{df})), however
additional features observed in our data are not seen in the measured dHvA
spectra and, conversely, we do not see the strongest peaks of the dHvA
oscillations. Instead, we see all SdH peaks and resolve additional features.
In addition, deep inside magnetically ordered state and in high fields $\mu$
is close to unity. The oscillatory part is a response to very small excitation
field superimposed on very large DC field. To further explore applicability of
the developed technique and rule out unexpected nonlinear or magnetic effects,
we also measured CeAgSb$_{2}$, which is isostructural to SmAgSb$_{2}$, but
showed a well-resolved, single frequency SdH oscillation of $2.5$ MG
\cite{myers2}. Figure \ref{CeOsc} shows the result of our measurements at
various temperatures below $T_{c}$. Clearly, we observe a single frequency
oscillations at 0.25 MG.%

%TCIMACRO{\FRAME{ftbphFU}{7.7607cm}{9.7838cm}{0pt}{\Qcb{Shubnikov - de Haas
%oscillations (raw data in terms of the resonant frequency shift) in single
%crystal CeAgSb$_{2}$ measured at indicated temperatures. The curves are offset
%for clarity.}}{\Qlb{CeOsc}}{ceosc.eps}{\special{ language "Scientific Word";
%type "GRAPHIC";  maintain-aspect-ratio TRUE;  display "ICON";
%valid_file "F";  width 7.7607cm;  height 9.7838cm;  depth 0pt;
%original-width 3.0277in;  original-height 3.8242in;  cropleft "0";
%croptop "1";  cropright "1";  cropbottom "0";
%filename 'CeOsc.eps';file-properties "XNPEU";}}}%
%BeginExpansion
%EndExpansion
A typical smallest oscillation amplitude of measured resonant frequency was
about $20$ Hz (Figure \ref{CeOsc}) whereas our resolution is better than $0.1$
Hz. We therefore are able to measure samples at least $10-100$ times smaller
than used here (depending on their resistivity and surface quality).

\begin{figure}
[ptbh]
\begin{center}
\includegraphics[
height=9.7157cm,
width=7.7607cm
]%
{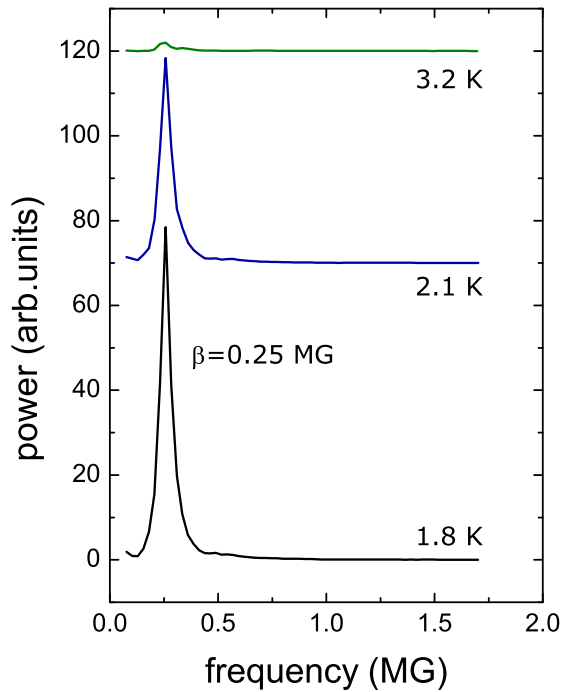}%
\caption{Power spectra of Shubnikov - de Haas oscillations observed in
CeAgSb$_{2}$ single crystal. The curves are offset for clarity.}%
\label{CePS}%
\end{center}
\end{figure}

Figure \ref{CePS} shows power spectra for the SdH oscillations, which is
similar to these reported previously \cite{myers2,jobiliong}. The amplitude of
oscillations decreased with the increasing temperature much more rapidly
compared to SmAgSb$_{2}$. This decrease, however, is very similar to recent
measurements of SdH effect in the same compound where signal vanishes just
above $3$ K \cite{jobiliong}.%

%TCIMACRO{\FRAME{ftbphFU}{7.7607cm}{9.7157cm}{0pt}{\Qcb{Power spectra of
%Shubnikov - de Haas oscillations observed in CeAgSb$_{2}$ single crystal. The
%curves are offset for clarity.}}{\Qlb{CePS}}{ceps.eps}%
%{\special{ language "Scientific Word";  type "GRAPHIC";
%maintain-aspect-ratio TRUE;  display "ICON";  valid_file "F";
%width 7.7607cm;  height 9.7157cm;  depth 0pt;  original-width 3.0277in;
%original-height 3.7965in;  cropleft "0";  croptop "1";  cropright "1";
%cropbottom "0";  filename 'CePS.eps';file-properties "XNPEU";}}}%
%BeginExpansion
%EndExpansion

\section{Discussion}

The difference in relative amplitudes between dHvA and SdH oscillations is not
surprising and is determined by the effective mass via $\alpha_{r}%
/M_{r}\propto m_{c}^{2}$ (see Eqs.\ref{dHvA} and \ref{SdH}). The temperature
dependence of the amplitude of the SdH oscillations in CeAgSb$_{2}$,
Fig.\ref{CeOsc}, allows us to roughly estimate the effective mass associated
with this orbit to be $m^{\ast}=0.6\pm0.15m_{0}$. From Fig. \ref{SmAmpT} an
estimate of the effective mass of the new orbit in SmAgSb$_{2}$ below $T_{N}$,
between $2$ K and $6$ K, gives an approximate value of $0.15m_{0}$, close to
the effective mass of the neighboring $\beta$ orbit evaluated in a similar
temperature/field range and to the value for a $\beta$ orbit in LaAgSb$_{2}$
\cite{myers2}. This low value also explains the substantial difference between
amplitudes of SdH and dHvA effects.

The results of electronic structure calculations of the two nonmagnetic (NM)
compounds from the family RAgSb$_{2}$ where R=Y and La were discussed
previously by Myers \textit{et al.}~\cite{myers2}. It was assumed that
substitution of Y or La for rare-earth element should not significantly change
the electronic structure near the Fermi energy, $E_{F}$, because 4\textit{f}
electrons are strongly localized. Observation of additional frequencies in
SmAgSb$_{2}$ was explained by the smaller residual resistivity and possible
change of the Fermi surface produced by the new periodicity due to
antiferromangetic ordering~\cite{myers2}. In this paper we revisit that
assumption and calculate the bandstructure specifically for the SmAgSb$_{2}$
using the tight-binding, linear muffin-tin orbital (TBLMTO) method within the
atomic sphere approximation (ASA)~\cite{TBLMTO1, TBLMTO2}. The local density
approximation parametrization due to von Barth and Hedin~\cite{BHXC} of the
spin-density functional has been used and the $4f$ states were treated as core states.

Since the spin structure in the AFM ordered state has not been experimentally
determined, we assume the simplest configuration with magnetic moments of the
two Sm$^{3+}$ ions (in $4f^{5}$ configuration) in the unit cell pointing in
the opposite directions. Such a configuration has an energy lower by $\Delta
E=84$ K/cell compared to the ferromagnetically ordered state. In order to
estimate the Ne\'{e}l temperature, $T_{N}$, we can use the Heisenberg model.
The $T_{N}$ calculated in the mean-field approximation is equal to $\frac
{2}{3}J_{0}$, where $J_{0}$ is the effective exchange parameter and it
corresponds to the sum of the interactions of one magnetic moment with all
others. Using the information about $\Delta E$ and the assumption that the
only nonzero interaction corresponds to the nearest neighbors, we obtain the
expression for $J_{0}=\Delta E/4$. The estimated $T_{N}=14$ K is in reasonable
agreement with Ne\'{e}l temperature of 9.5 K observed in the compound.%

%TCIMACRO{\FRAME{ftbpFU}{9.1028cm}{6.4976cm}{0pt}{\Qcb{(color online) Band
%structure of SmAgSb$_{2}$ calculated from ASA TBLMTO. $E_{F}$ corresponds to
%zero energy. The nonmagnetic result is shown by the solid lines; the
%antiferromagnetic case is shown by the dotted lines.}}{\Qlb{bndstr}}%
%{bnd.eps}{\special{ language "Scientific Word";  type "GRAPHIC";
%maintain-aspect-ratio TRUE;  display "ICON";  valid_file "F";
%width 9.1028cm;  height 6.4976cm;  depth 0pt;  original-width 5.7519in;
%original-height 4.0932in;  cropleft "0";  croptop "1";  cropright "1";
%cropbottom "0";  filename 'bnd.eps';file-properties "XNPEU";}}}%
%BeginExpansion
\begin{figure}
[ptb]
\begin{center}
\includegraphics[
height=6.4976cm,
width=9.1028cm
]%
{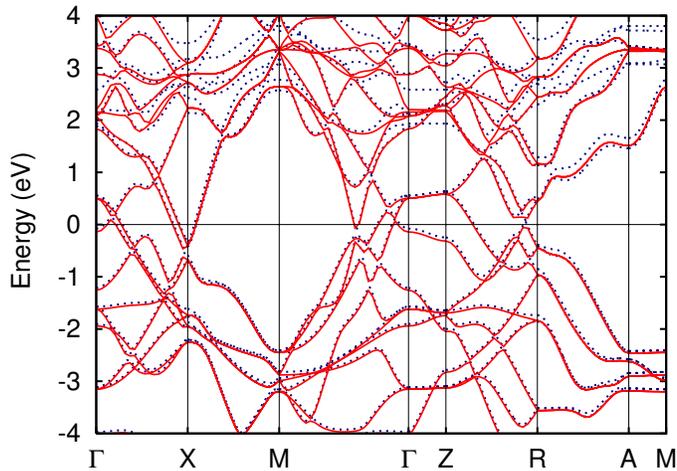}%
\caption{(color online) Band structure of SmAgSb$_{2}$ calculated from ASA
TBLMTO. $E_{F}$ corresponds to zero energy. The nonmagnetic result is shown by
the solid lines; the antiferromagnetic case is shown by the dotted lines.}%
\label{bndstr}%
\end{center}
\end{figure}
%EndExpansion

\begin{figure}
[ptbh]
\begin{center}
\includegraphics[
height=12.8546cm,
width=7.0973cm
]%
{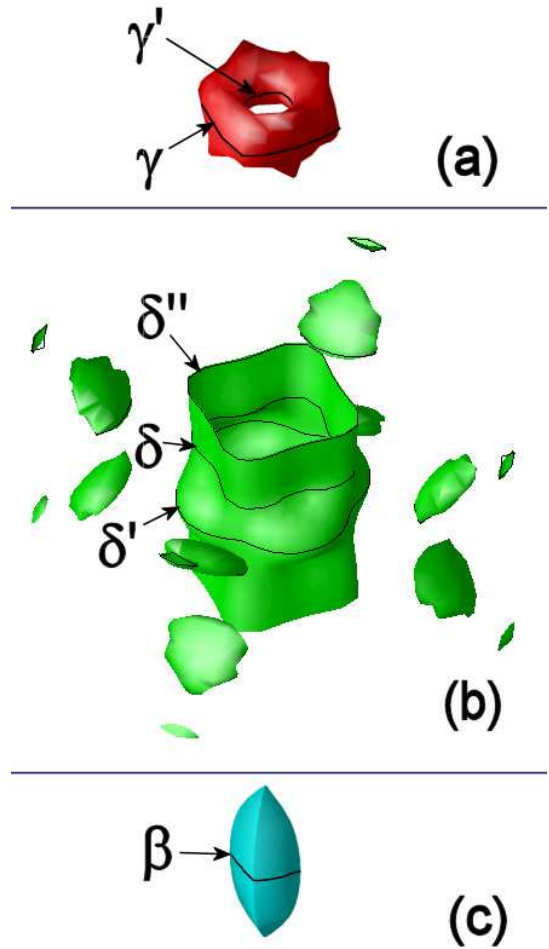}%
\caption{ Fermi surface of SmAgSb$_{2}$ corresponding to (a) band 1, (b) band
2 and (c) band 3. The extremal orbits indicated by greek letters and shown by
arrows.}%
\label{banda}%
\end{center}
\end{figure}

Figure~\ref{bndstr} shows the electronic structure of SmAgSb$_{2}$ along
several high-symmetry directions. The assumed AFM ordering (bands shown by
dashed lines) lifts the degeneracy in some symmetry directions. However, such
modification of the electronic structure does occur for the bands crossing
Fermi level. Hence, almost all the difference between the Fermi surfaces of
antiferromagnetic SmAgSb$_{2}$ and nonmagnetic YAgSb$_{2}$~\cite{myers2} is a
downshift of the Fermi level position compared to the band labeled $"1"$ in
Ref.~\cite{myers2}. This shift of $E_{F}$ leads to the modification of nearly
spherical FS sheet (band $1$) centered at a $\Gamma$ point, which is
transformed into a torus as shown in Fig.~\ref{banda} (a) with frequency of
$1$ MG for $H||c$-axis for the internal circle orbit $\gamma^{\prime}$, close
to the frequency of $0.82$ MG of an orbit labeled $\beta$ in Figure
(\ref{SmPST05}) of the second band.%

%TCIMACRO{\FRAME{ftbphFU}{7.0973cm}{12.8546cm}{0pt}{\Qcb{ Fermi surface of
%SmAgSb$_{2}$ corresponding to (a) band 1, (b) band 2 and (c) band 3. The
%extremal orbits indicated by greek letters and shown by arrows.}}{\Qlb{banda}%
%}{banda.eps}{\special{ language "Scientific Word";  type "GRAPHIC";
%maintain-aspect-ratio TRUE;  display "ICON";  valid_file "F";
%width 7.0973cm;  height 12.8546cm;  depth 0pt;  original-width 2.2035in;
%original-height 4.0075in;  cropleft "0";  croptop "1";  cropright "1";
%cropbottom "0";  filename 'banda.eps';file-properties "XNPEU";}}}%
%BeginExpansion
%EndExpansion

Another FS sheet which changes significantly corresponds to band $2$ and is
shown in Fig.~\ref{banda} (b). Band $2$ is also centered at a $\Gamma$ point.
In addition to mostly cylindrical part with axis along $k_{z}$ as in
YAgSb$_{2}$, eight pillow-shaped segments now appear around the main cylinder
with the frequency of the $\delta"$ orbit equal to $1.78$ MG. We therefore
conclude that previously unresolved peak (closest to $\beta$) in SmAgSb$_{2}$
is determined by the $\gamma^{\prime}$ orbit. The appearance of this orbit is
caused by the shift of $E_{F}$ position with respect to band $1$ compared to
YAgSb$_{2}$ where this orbit is absent. This shift is produced by filling $4f$
states in Sm$^{3+}$ by five electrons that lead to a slight change of the
d-band occupancy compared to Y-based compound.

In conclusion, a high resolution method for contact-less measurements of the
resistivity via normal skin depth was developed to probe Shubnikov-de Haas
oscillations in small metallic samples. Torque measurements can be used to
obtain de Haas- van Alphen oscillations and presented technique can be applied
to the same small samples to obtain Shubnikov-de Haas signal without modifying
the samples (to attach contacts). The application of the method was
demonstrated on RAgSb$_{2}$ system. Fine details of the oscillation spectrum
(a new $\gamma^{\prime}$ orbit) was resolved experimentally and explained with
refined bandstructure calculations.

\begin{acknowledgments}
We thank Bruce N. Harmon for useful discussions. Ames Laboratory is operated
for the U.S. Department of Energy by Iowa State University under Contract No.
W-7405-ENG-82. This work was supported in part by the Director for Energy
Research, Office of Basic Energy Sciences. R.P. acknowledges support from the
Alfred P. Sloan Foundation.
\end{acknowledgments}


\begin{thebibliography}{99}                                                                                               %


\bibitem {shoenberg}D. Shoenberg, \textit{\textquotedblright Magnetic
oscillations in metals\textquotedblright}, Cambridge University Press (1984).

\bibitem {kartsovnik}M. V. Kartsovnik, Chem. Rev. \textbf{104}, 5737-5781 (2004).

\bibitem {jackson}J. D. Jackson, Classical Electrodynamics (John Wiley \&
Sons, Inc., New York, 1998), p. 220.

\bibitem {pippard}A. B. Pippard, Phil. Tranc. Roy. Soc. \textbf{A250}, 325 (1957).

\bibitem {prozorov}R. Prozorov, R. W. Giannetta, A. Carrington, and F. M.
Araujo-Moreira, Phys. Rev. B \textbf{62}, 115-118 (2000).

\bibitem {prozorov2}R. Prozorov, R. W. Giannetta, A. Carrington, P. Fournier,
R. L. Greene, P. Guptasarma, D. G. Hinks, and A. R. Banks, Appl. Phys. Lett.
\textbf{77}, 4202-4204 (2000).

\bibitem {choi}E. S. Choi, E. Jobilong, A. Wade, E. Goetz, J. S. Brooks, J.
Yamada, T. Mizutani, T. Kinoshita, and M. Tokumoto, Phys. Rev. B \textbf{67},
174511/(1-8) (2003).

\bibitem {coffey}T. Coffey, Z. Bayindir, J. F. DeCarolis, M. Bennett, G.
Esper, and C. C. Agosta, Rev. Sci. Inst. \textbf{71}, 4600-4606 (2000).

\bibitem {cooper}J. R. Cooper, Phys. Rev. B \textbf{54}, R3753 (1996).

\bibitem {brylak}M. Brylak, M. H. Moeller, and W. Jeitschko, J. Sol. Stat.
Chem. \textbf{115}, 305 (1995).

\bibitem {sologub}O. Sologub, H. Noeel, A. Leithe-Jasper, P. Rogl, and O. I.
Bodak, J. Sol. St. Chem. \textbf{115}, 441 (1995).

\bibitem {myers1}K. D. Myers, S. L. Bud'ko, I. R. Fisher, Z. Islam, H.
Kleinke, A. H. Lacerda, and P. C. Canfield, J. Mag. Mag. Mater. \textbf{205},
27 (1999).

\bibitem {myers2}K. D. Myers, S. L. Bud'ko, V. P. Antropov, B. N. Harmon, P.
C. Canfield, and A. H. Lacerda, Phys. Rev. B \textbf{60}, 13371 (1999).

\bibitem {jobiliong}E. Jobiliong, J. S. Brooks, E. S. Choi, H. Lee, and Z.
Fisk, Phys. Rev. B \textbf{72}, 104428/1 (2005).

\bibitem {canfield}P. C. Canfield and Z. Fisk, Philos. Mag. B \textbf{65},
1117 (1992).

\bibitem {TBLMTO1}O. K. Andersen, Phys. Rev. B \textbf{12}, 3060 (1975).

\bibitem {TBLMTO2}O. K. Andersen, O. Jepsen, Phys. Rev. Lett \textbf{53}, 2571 (1984).

\bibitem {BHXC}U. von Barth and L. Hedin, J. Phys. C \textbf{5}, 1629 (1972).
\end{thebibliography}
\end{document}